\documentclass[slac_one]{revtex4}
\usepackage{graphicx}
\usepackage{fancyhdr}
\def\lapp{\mathrel{\rlap{\raise.5ex\hbox{$<$}}
                    {\lower.5ex\hbox{$\sim$}}}}
\def\gapp{\mathrel{\rlap{\raise.5ex\hbox{$>$}}
                    {\lower.5ex\hbox{$\sim$}}}}
\newcommand{\sm}{SM}

\begin{document}

\title{Nonstandard, strongly interacting  spin one $t \bar t$ resonances.} %% Paper title goes here


\author{Rohini M. Godbole}
\affiliation{Centre for High Energy Physics, Indian Institute of Science, Bangalore 560012, India.}
\author{Debajyoti Choudhury} 
\affiliation{Dept. of Physics and Astrophysics, University of Delhi, Delhi 110007, India.}

\begin{abstract}
Examining theories with an extended 
strong interaction sector such as axigluons
or flavour universal colorons, we find that the constraints obtained from 
the current data on $t \bar t$ production at the Tevatron are  in the 
range of $\sim {\cal O}$ TeV and thus competitive with those obtained 
from the dijet data. 
We point out that for large axigluon/coloron masses, 
the limits on the coloron mass
may  be different than those for the axigluon
even  for $\cot \xi = 1$.
We also compute the
expected forward-backward  asymmetry for the case of the axigluons which
would allow it to be discriminated against the SM as also the colorons. We
further find that at the LHC, the signal should be visible in the $t \bar t$
invariant mass spectrum for a wide range of axigluon and coloron masses
that are still allowed.  We point out how top polarisation
may be used to  further discriminate the axigluon and coloron case from the SM
as well as from each other.

\end{abstract}

\maketitle

\thispagestyle{fancy}


\section{INTRODUCTION} % Section title should be in all capitals.
The importance of the study of top quark physics at the current stage 
in Particle Physics can hardly be overemphasized. Apart from 
its crucial role in the test of the Standard Model (SM)  at the 
loop level, the closeness of the top mass to the Electroweak Symmetry 
Breaking (EWSB) scale accord it a special role 
in virtually any alternative to the Higgs mechanism. 
Thus, the 
production of top quarks  at the colliders can be a {\bf low} energy probe
of the {\bf high} scale physics that might be triggering the EWSB. 
Already at the Tevatron, this is a topic
of much attention~\cite{Cabrera:2006ya,Lannon:2006vm} and  a
top factory such as the LHC would provide valuable information on the \sm\
as well as physics beyond it~\cite{Beneke:2000hk}.

In our work~\cite{plbcgsw},  we revisit the 
issue of strongly interacting spin one  gauge bosons and 
their contribution to 
$t \bar t$ production at hadronic colliders. We consider two classes of 
models : 1) Flavour universal colorons  which are  present in theories of 
extended color gauge theories and 2) Axigluons which exist in theories of 
chiral colour. Although neither of these have preferentially larger 
couplings to the $t \bar t$ pair, unlike  Kaluza 
Klein gluons~\cite{Guchait:2007ux} or extended technicolour models,
we demonstrate that even the current data  on $t \bar t $ 
production yield very competitive 
constraints on the masses and coupling of 
these gauge bosons.

\section{Axigluon and Flavour Universal Coloron Models}
Arising in 
unifiable models of chiral colour~\cite{Frampton:1987dn,Frampton:1987ut},  
Axigluons are massive, strongly interacting gauge bosons 
with an axial vector 
coupling ${\frac {1}{2}} g_s \gamma_\mu \gamma_5 \lambda^a$,  where 
$\lambda_a$ are the usual Gell-Mann matrices. 
In the simplest models, a high scale strong interaction gauge group 
of $SU(3)_L \times SU(3)_R$ is broken to the familiar 
$SU(3)_c \equiv SU(3)_{L+R}$,
 resulting in massive states with the aforementioned coupling. 
This carries through for all generalisations of chiral color. Embedding this 
in a unified group implies $m_A \sim 250$ GeV and
hence was searched for very actively at the Tevatron.

Flavour universal Colorons~\cite{Chivukula:1996yr} arise in
 models with an extended colour gauge group. The latter 
were part of the 
general effort to understand the mechanism of EW symmetry breaking and the 
large mass of the top in the same framework. With a top quark 
condensate enhancing $m_t$ as well as  driving EWSB, 
specific examples of this idea are topcolor~\cite{Hill:1991at,Hill:1993hs} 
and topcolor assisted technicolor~\cite{Hill:1994hp}. The colour group at the
high scale is $SU(3)_I \times SU(3)_{II}$---both being vector-like---
which then breaks to $SU(3)_c$ 
giving rise to the massive `colorons'.
Variants of the model essentially differ in the way 
generations couple to the colorons. The one we consider is the
simplest and is characeterised by a 
universal  coupling (${\frac {1}{2}} g_s \cot \xi  \gamma_\mu \lambda^a$)
to {\it all}  the quarks. These models can be 
be grafted into a single Higgs doublet model and has a {\bf naturally}
heavy top. Understandably, EW precision measurements 
restrict the model in the mass-coupling ($M_C$--$\cot \xi$)
plane.

\section{PHENOMENOLOGY AT THE TEVATRON}
The broad, strongly interacting axigluon and coloron resonances, $A,C$, can be
copiously produced at a hadronic collider and
thus could show up as an additional, resonant contribution to the dijet 
production cross-section 
($q \bar q \rightarrow A^* (C^*) \rightarrow q' \bar q'$). 
At the Tevatron, the dominance of the $q \bar q$ flux implies
that these
contributions can be quite large. 
At the time of writing the paper~\cite{plbcgsw}, the best available limits 
on the axigluon and coloron masses came from the 
dijet sample~\cite{Abe:1997hm,Giordani:2003ib,Yao:2006px} 
which 
rules out an axigluon of mass less\footnote{At this conference
newer limits have been quoted. These are not included in this report.}
than $980$ GeV.
The same limit is quoted for coloron for $\cot \xi =1$ for the flavour 
universal case. In the approximation of neglecting the width and 
interference with background, the limit on coloron masses can get only 
stricter with increasing $\cot \xi$. 

Note, though, that the axigluon and coloron cases differ in a
crucial manner: while the $s$-channel coloron exchange amplitude 
can interfere with a similar QCD amplitude (for simplicity, let 
us consider $q \neq q'$), this is nonexistent for the axigluon.
For small masses, the resonance is narrow;
with the difference 
being negligible, the limits for axigluon and coloron
with $\cot \xi =1$ would be nearly identical. However, as we will see
shortly, the approximation may not be
justified for higher mass resonances and it would be interesting to 
examine how the limits obtained from dijet analysis are affected.

In this work, we are interested 
in $t \bar t$ production.
 At the tree level, the presence of either $A$ or  
$C$ can affect $t \bar t$ production only as far as the 
$q\bar q$-initiated subprocess is concerned, leaving the $gg$-initiated 
subprocess unaltered.  
We refrain from reproducing 
the expressions for differential cross-sections 
% including the interference effects between the SM and the coloron
% amplitude. These 
which are available in Ref.~\cite{plbcgsw}.

%%%
\begin{figure}[!h]
\vspace*{-28ex}
\includegraphics*[scale=0.53]{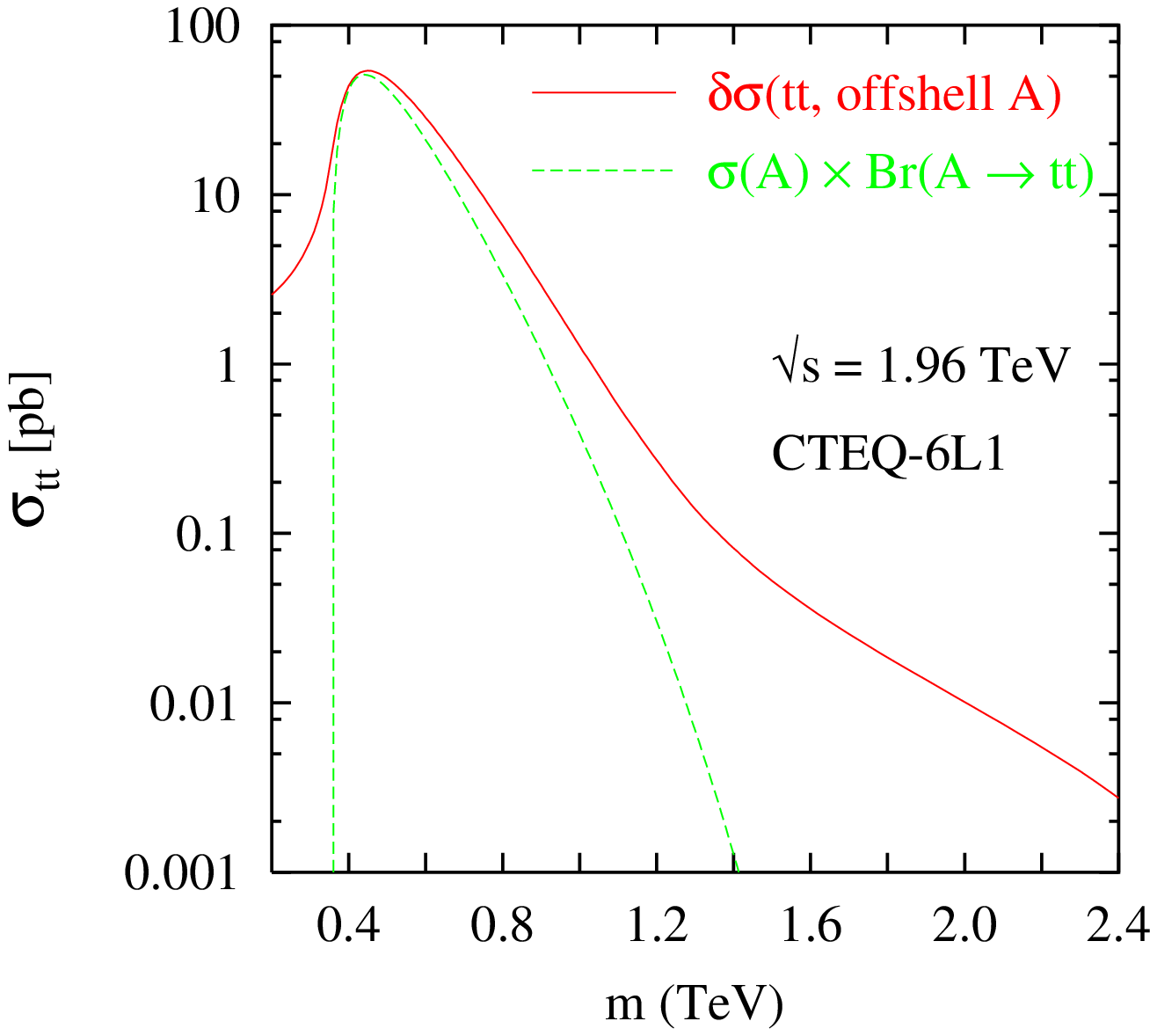}% {../ichep08_talk/br_sig.ps}
\hspace*{5ex}
\includegraphics*[scale=0.4]{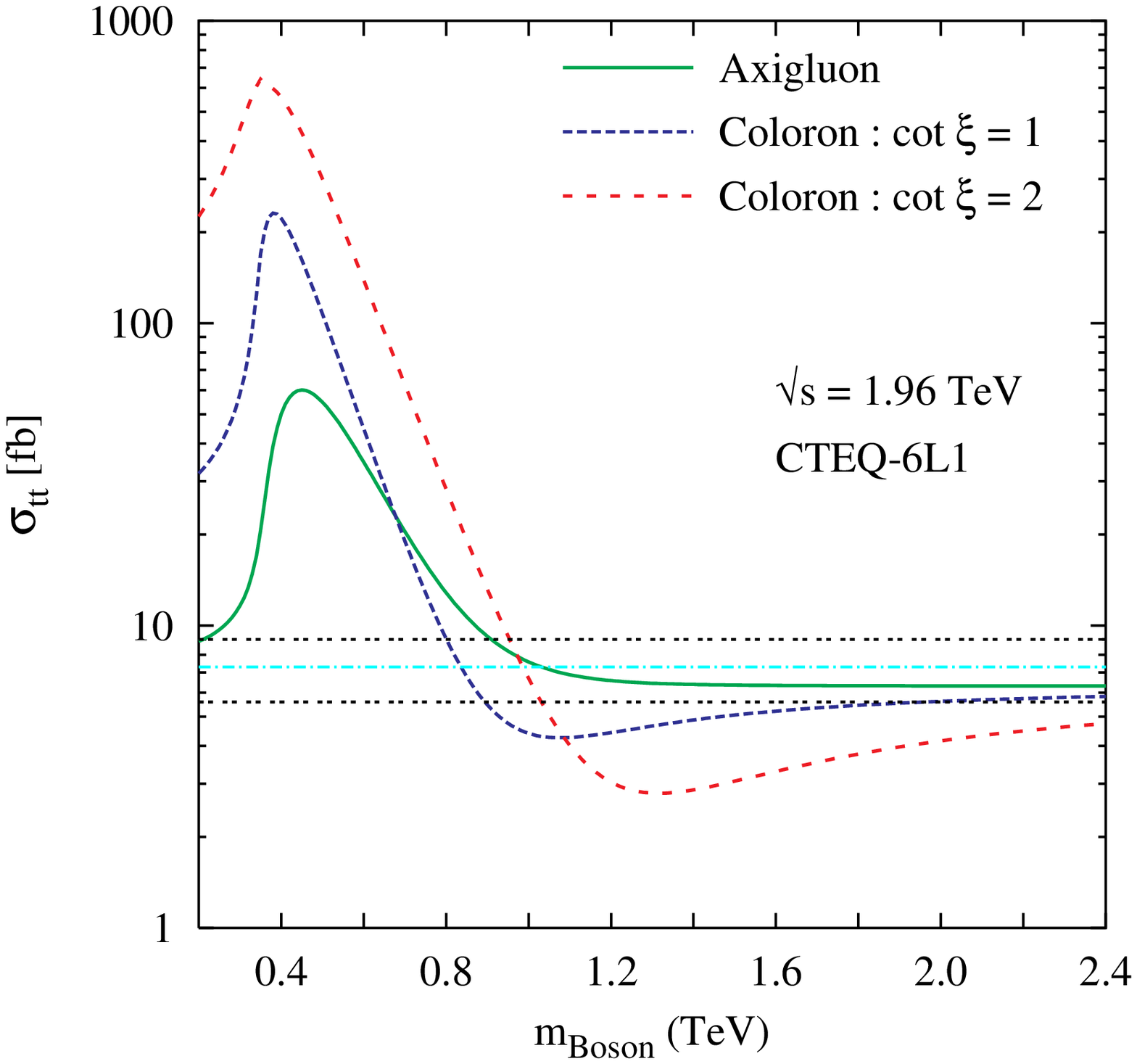}%{../ichep08_talk/cs_tev_b.ps}
\caption{
(a) A comparison of the deviation of the total $t \bar t$ cross section
caused by the presence of an axigluon (solid line) with the
resonant production followed by decay.
(b) $\sigma(t \bar t)$  at the Tevatron as a function
of the axigluon (coloron) mass. The solid (green) line
corresponds to the axigluon case. The short- (blue) and long-dashed (red)
lines correspond to the flavour universal coloron for $\cot \xi = 1 \, (2)$ 
 respectively. The horizontal lines correspond
to the CDF central value
and the $95 \% $  confidence level band~\cite{Cabrera:2006ya}.
CTEQ-6L1 parton distributions evaluated at $Q = m_t$
were used alongwith  the 
appropriate $K$--factor~\protect\cite{Campbell:2006wx}.
}
\label{widths_cs:f1}
\end{figure}
%%%
The widths are substantial for either of 
$A/C$ capable of decaying into a top-pair. 
Furthermore, the partial widths into a top-pair are 
different even for $\cot\xi = 1$. 
The large widths 
imply that the (narrow-width) approximation of resonant production and
subsequent decay is no longer a good one. This is borne out by
Fig.\ref{widths_cs:f1}$a$, wherein we compare the narrow-width contribution 
to $t \bar t$ production, viz. $\sigma (A) \times BR (A \rightarrow t \bar t)$
with the exact result, namely
 $\delta \sigma \equiv \sigma_A (t \bar t) - \sigma_{SM} (t \bar t)$.
%
% $\sigma(t \bar t)$
% at the Tevatron expected, for an axigluon, in the
% resonant (incoherent) approximation with the exact value of the
% deviation in $\sigma(t \bar t)$ obtained by integrating over the
% (large) width of the resonance. To be more specific, we compare
% $\sigma (A) \times BR (A \rightarrow t \bar t)$
% with $\delta \sigma \equiv \sigma_A (t \bar t) - \sigma_{SM} (t \bar t)$.
%
The effect is indeed substantial. For the dijet case
the effect will be smaller, but may still be non-negligible and hence
might affect the limits on axigluon/coloron masses obtained from the 
dijet data. Further, it is to be noted that the axial coupling of 
the $A$ gives rise to a forward-backward asymmetry for the $t$ 
quark~\cite{Sehgal:1987wi}. For colorons, of course,
no such asymmetry can exist. 

\begin{figure}[tbh]
%\label{exclusion:f3}
\vspace*{-10ex}
\includegraphics*[scale=0.38]{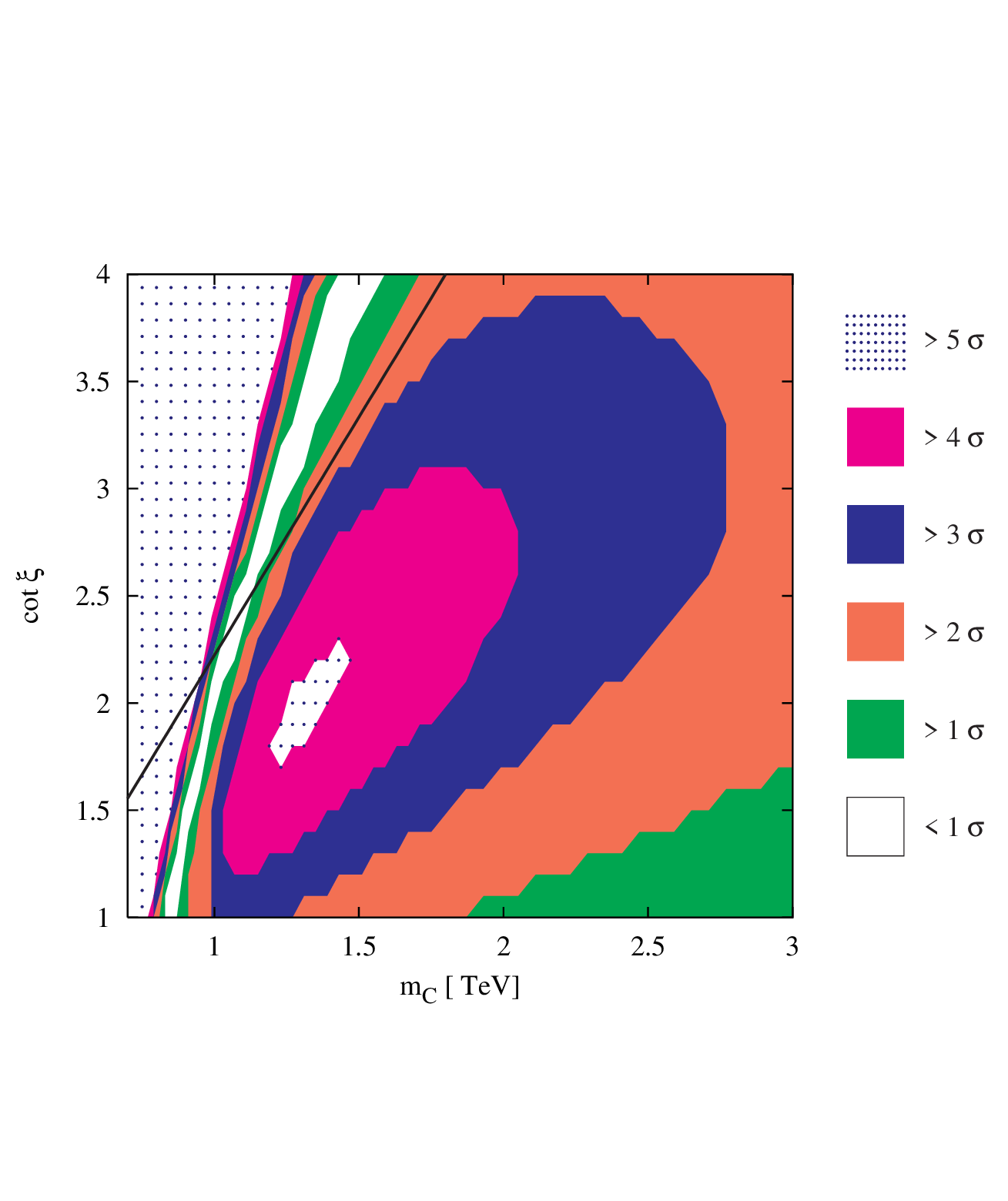}%{../ichep08_talk/color_cont_dots_new.ps}
\hspace*{5ex}
\includegraphics*[scale=0.50]{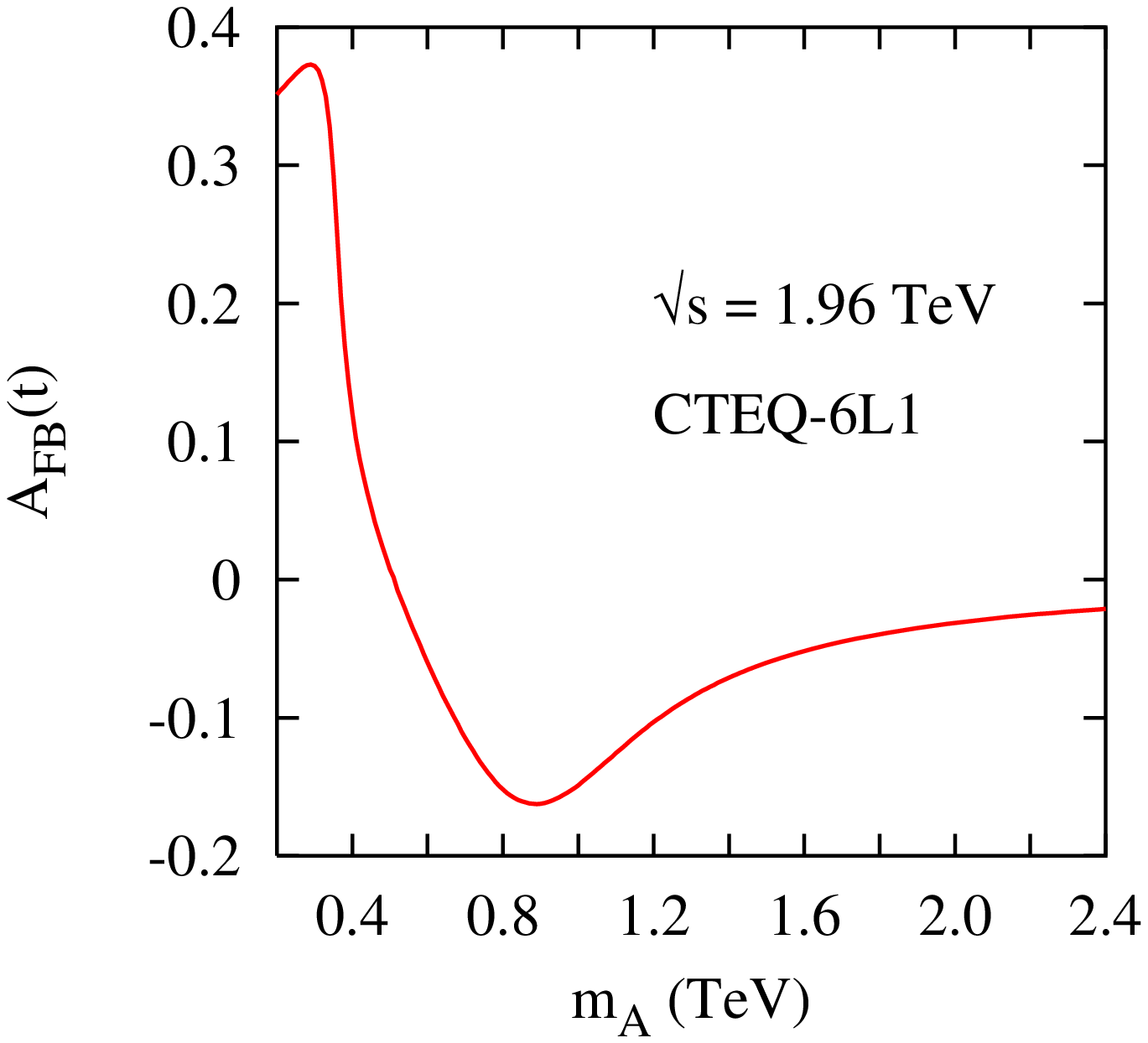}% {../ichep08_talk/asym_tev.ps}
\vspace*{-13ex}
\caption{Left panel gives the exclusion region in the $\cot \xi$ -- $m_C$ plane using the
$t \bar t$ data at Tevatron~\protect\cite{Cabrera:2006ya}.
The solid curve shows the constraint imposed by the
$\rho$ parameter $m_C/\cot \xi \gapp 450$.\label{exclusion:f3} 
The right panel shows the FB asymmetry in $t\bar t$ production at the 
LHC as a function of the axigluon mass}
\end{figure}

%%%
In Figure~\ref{widths_cs:f1}$b$, the results of the $t \bar t$ cross-sections
expected fot the coloron/axigluons at the Tevatron are shown as a function
of the mass of the boson and for different values of $\cot \xi$.
A few facts are to be noted.
\begin{itemize}
\item For axigluon case due to the different parity of the SM amplitude and 
the axigluon amplitude, the  interference term does not contribute to the 
total rate.\\[-5ex]
\item For coloron the intereference term contributes and also changes sign as
$q \bar q$ subprocess energy passes through $M_C$, depending on $\cot \xi$.
\\[-5ex]
\item For masses of massive gluon above $2 m_t$ not just the inteference
term but  the squared contribution of the new amplitude are different for 
coloron and axigluon.
\end{itemize}
The data indicated by the horizontal lines  in Figure~\ref{widths_cs:f1}$b$
 and taken
from ~\cite{Cabrera:2006ya}  corresponds to:
\[
\sigma(p + \bar p \to t + \bar t + X; \sqrt{s} = 1.96 \, {\rm TeV})
 = 7.3 \pm 0.5 \, {(stat)} \pm 0.6 \, {(syst)} \pm 0.4 \, {(lum)} \ {\rm pb}.
\]
Using these, we get for the axigluon $M_A > 910$ GeV at $95 \%$ C.L., whereas
for the coloron, for $\cot \xi =1$, $800 < M_C < 895$ {\bf and} $M_C > 1960$ 
are allowed  at the same C.L. These limits are quite competitive with those
available from the dijet analysis and are in fact different for the coloron
and the axigluon even for $\cot \xi =1$. Furthermore, the coloron mass limits
depend on $\cot \xi$ non-monotonically (a consequence of the interference 
term), as is evident in both
Figs.~\ref{widths_cs:f1}$b$\&\ref{exclusion:f3}$a$, the second 
%%%
%%%
displaying the exclusion region for the coloron
in the $\cot \xi$--$m_C$ plane. Note that
the consistency of certain
regions in the paramter space with the data cannot be interpreted
as evidence for the colorons as 
the same data are consistent with the \sm\ as well.  

The parity violating axigluon coupling would also lead to a 
forward-backward asymmetry at the Tevatron as 
shown in Figure~\ref{exclusion:f3}$b$. Two things
are to be noted here. Our calculation corrects a mistake in Ref.~\cite{Sehgal:1987wi}. Secondly for the masses accessible at the Tevatron these are quite 
sizable and substantially larger than the the one expected due to QCD radiative
corrections~\cite{kuhn}. This agrees with the detailed comparisons
of the latter with those expected for  axigluon contribution  performed in 
Ref. \cite{kuhnlatest},  which appeared soon after our work. In fact the 
asymmetry caused by the axigluon resonance will have a different  dependence on the phase space variabels from those caused by QCD effects. With this, one 
could in fact use these asymmetries (or absence thereof) to obtain constraints 
on the axigluons.
%%%
\section{PHENOMENOLOGY AT THE  LHC}
\begin{figure}[tbh]
\vspace*{-5ex}
\includegraphics*[scale=0.3]{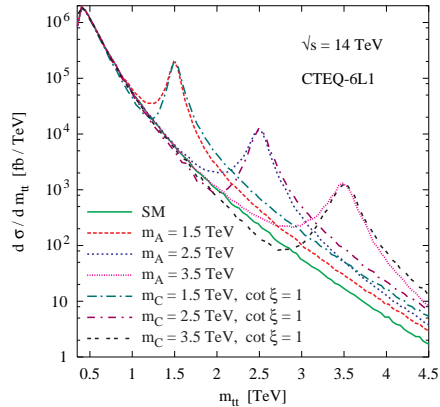}%{../ichep08_talk/mtt_lhc.ps}
\vspace*{-3ex}
\caption{The expected $m_{t \bar t}$
spectrum at the LHC in presence of either axigluons or colorons of
a specified mass, along with the SM expectations.\label{asymmetrylhc:f4} 
}
\end{figure}

With the 
Tevatron pushing the limits on the axigluon and coloron masses higher,
it is natural to investigate the prospects  
at the LHC. As gluon fluxes would dominate over $q \bar q$,
it is imperative to 
look at differential distributions, in particular that 
in the invariant mass of the $t \bar t$ pair. We see from the right panel of
Figure\ref{asymmetrylhc:f4} that, for the first peak,  assuming even only a 
$10 \%$ efficiency,  there will be about $\sim 10^4$ events with $10$ fb$^{-1}$
and thus a good chance of being able to see them at the LHC. For such masses, 
the effect of $m_t$ on the decay width is negligible and, for $\cot \xi =1$, 
the differential cross sections are virtually the same at the 
resonance.

Unfortunately, LHC being a $pp$ machine, no FB asymmetry can be constructed
for the axigluon case. However, the correlation between  helicities of 
$t$ and $\bar t$ carry the information on the heavy gluon contribution. 
Instead one can construct 
$
{\cal R}_\Delta(m_{tt}) \equiv
    \left[ \int_{m_{tt} - \Delta}^{m_{tt} + \Delta} \, d m_{tt} \,
                    \frac{d \sigma_{-}}{d \, m_{tt}} \right]
    \;
    \left[ \int_{m_{tt} - \Delta}^{m_{tt} + \Delta} \, d m_{tt} \,
                    \frac{d \sigma_{+}}{d \, m_{tt}}  \right]^{-1},
$
where $\sigma_{\pm}$ refer to the cross sections for the
product of the $t$ and $\bar t$ helicities to be $\pm 1$ respectively.
These are, in essence, like the spin-spin correlation measurements which 
have been suggested for the study of $CP$/spin  properties of a resonance 
which can decay into a $t \bar t$~\cite{atlas}. In fact, this
can then provide an additional handle to distinguish between the two cases at
hand.

\begin{figure}[tbh]
%\label{lrasymm:f5}
\includegraphics*[scale=0.3]{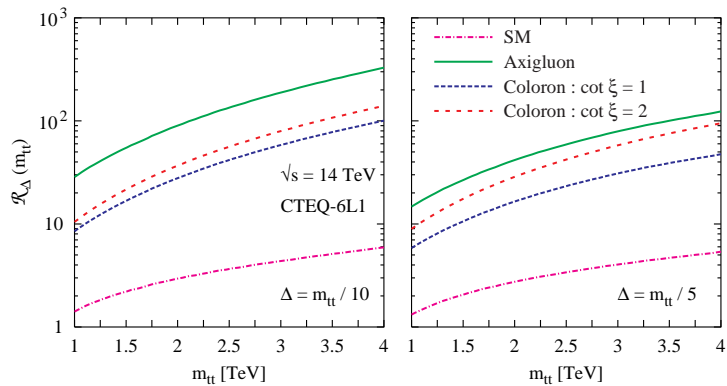}% {../ichep08_talk/lr_lhc.ps}
\caption{The ratio of the partial cross-sections
${\cal R}_\Delta(m_{tt} = m_{\rm Boson})$
as a function of the boson mass. The
two panels correspond to different values of $\Delta$
.\label{lrasym:f5}}
\end{figure}

\section{CONCLUSIONS}
In conclusion, constraints on the axigluons and colorons obtained
from $t \bar t$ production at the Tevatron are indeed competetive with those 
from dijets.  The forward-backward  asymmetry at the Tevatron can help 
constrain the axigluon further. Nature of interference term with the SM 
amplitude are different for axigluon and coloron cases. The limits obtained 
from $t \bar t$ production on coloron masses depend on $\cot \xi$ non 
monotonically.  The zero width approximation too crude at larger
masses and mass limits obtained for dijets for coloron may not be the
same as that of an axigluon, even for $\cot \xi = 1$.
At the LHC differential distribution in $m_{t \bar t}$ can
show up evidence for colorons and axigluons. Their effect is measurable. 
Further, a variable similar to the spin spin correlations can help 
distinguish between the two further.

\begin{acknowledgments}
This work was in part supported under the Indo French
Centre for Promotion of Advanced Research Project 3004-B
and project number
SR/S2/RFHEP-05/2006 of the Department of Science and Technology (DST), India,
as  well as a grant of J.C. Bose Fellowship to R.G. from the DST.
\end{acknowledgments}

\end{document}